\documentclass[useAMS,usenatbib,referee]{mn2e}

\usepackage{subfigure}

\usepackage{amsmath}
\usepackage{amssymb}
\usepackage{epsfig}
\usepackage{times}

\def\ggrav{\,{{\bf g}_{\rm grav}}}
\def\grad{\,{{\bf g}_{\rm rad}}}
\def\kms{\,{\rm km~s^{-1}}}

\def\Rstar{\,{\rm R}_\star}

\def\vesc{\,{v_{\rm esc}}}

\def\Flux{\mathbf F}
\def\vel{{\mathbf u}}

\newcommand{\beq}{\begin{equation}}
\newcommand{\eeq}{\end{equation}}
\newcommand{\beqa}{\begin{eqnarray}}
\newcommand{\eeqa}{\end{eqnarray}}

\def\apj{\,{ApJ}}
\def\apjs{\,{ApJS}}

\title[Stellar winds above the photon-tiring limit]{On the behaviour of stellar winds that exceed the photon-tiring limit}
\author[A. J. van Marle et al.]
{Allard~Jan~van~Marle$^{1,2}$, Stanley~P.~Owocki$^1$ and Nir~J.~Shaviv$^3$ \\
$^1$Bartol Research Institute, University of Delaware, Newark, DE 19716, USA\\
$^2$Centre for Plasma Astrophysics, KU Leuven, Celestijnenlaan 200B, bus 2400, B-3001 Leuven, Belgium \\
$^3$Racah Institute of Physics, Hebrew University, Giv'at Ram, Jerusalem 91904 Israel
}
\date{Submitted 2008 August 1}

\pagerange{\pageref{firstpage}--\pageref{lastpage}}
\pubyear{2008}

\begin{document}

\maketitle

\label{firstpage}

\begin{abstract}
Stars can produce steady-state winds through radiative driving as long as
the mechanical luminosity of the wind does not exceed the radiative
luminosity at its base.  This upper bound on the
mass loss rate is known as the photon-tiring limit.  Once above this limit, the
radiation field is unable to lift all the material out of
the gravitational potential of the star, such that only part of it can
escape and reach infinity.  The rest stalls and falls back toward
the stellar surface, making a steady-state wind impossible.
Photon-tiring is not an issue for line-driven winds since they cannot achieve
sufficiently high mass loss rates. It can however become important if the star
exceeds the Eddington limit and continuum interaction becomes the
dominant driving mechanism.

This paper investigates the time-dependent behaviour of stellar
winds that exceed the photon-tiring limit, using 1-D numerical simulations
of a porosity moderated, continuum-driven stellar wind.  We find that
the regions close to the star show a hierarchical pattern of high
density shells moving back and forth, unable to escape the
gravitational potential of the star.  At larger distances, the flow
eventually becomes uniformly outward, though still quite variable.
Typically, these winds have a very high density but a terminal flow
speed well below the escape speed at the stellar surface.  Since most of the
radiative luminosity of the star is used to drive the stellar wind,
such stars would appear much dimmer than expected from the
super-Eddington energy generation at their core. 
The visible luminosity typically constitutes less then half of the total energy flow and can become 
as low as ten percent or less for those stars that exceed the photon-tiring
limit by a large margin.
\end{abstract}


\section{Introduction}

The high luminosity of hot, massive stars leads to strong
radiatively driven mass loss. 
In the more or less steady winds of O, B, and WR stars, 
the driving is through scattering of radiation by an ensemble of
spectral lines 
(e.g., Castor, Abbott, \& Klein 1975, hereafter \cite{cak}).
The tendency of these driving lines to become saturated with increasing
density acts as an effective wind regulator,
limiting the wind mass loss rates to ca.~$10^{-4} M_{\odot}$/yr in
even the most luminous stars.
While potentially quite important for the star's evolution, this level
of mass loss is energetically of only minor significance,
with a total (kinetic and potential) mechanical luminosity of only a
few percent of the ``photon tiring limit'' set by the total available
stellar luminosity.


By contrast, the mass loss rates during the giant eruptive phase of
Luminous Blue Variable (LBV) stars like $\eta$~Carinae
are inferred to be as high as $0.1-1 \, M_{\odot}$/yr,
much higher than can be reasonably
explained by models based on line-driving \citep{so06}.
In the case of $\eta$~Carinae,
the inferred average mechanical luminosity during 
its ca.\ decade-long eruption in the 1840's is comparable to the
estimated average radiative luminosity, 
$\sim 2.5 \times 10^{7}~L_{\odot}$, 
during this time.
This luminosity likely well exceeds the {\em Eddington limit}, at which 
the radiative force associated with {\em continuum} scattering by free
electrons exceeds the inward force of gravity, implying then that the 
extreme mass loss can be driven by continuum, instead of line,
opacity.

Unlike line driving, continuum driving does not suffer saturation at
high densities, meaning then it can {\em initiate} an {\em arbitrarily large}
mass flux, since all available photon-energy can be used to drive the
matter.
However, if the mechanical luminosity needed to lift this mass flux
fully out of the gravitational potential exceeds the stellar luminosity,
then the ``tiring'' or reduction in radiative energy flux will make it
unable to sustain the outflow.
This paper presents time-dependent hydrodynamical simulations of
the complex combination of stagnated outflow and subsequent infall
that occurs in continuum-driven models with initial, base flux that
exceeds this photon-tiring limit.


Previous analyses of continuum-driven mass loss have focussed on
developing steady-state wind outflow models with mass loss rates 
that approach, but 
do not exceed the photon-tiring limit.
To provide the necessary driving regulation and limitation of the mass flux, 
such models invoke a  ``porosity'' moderation of the opacity associated
with a clumped medium \citep{s98,s01}. 
 Such clumps can result from instabilities, such as the photon-bubble
instability \citep{s76,hak97, st99, b01, s01b} in the stellar atmosphere and the upper 
layers of the stellar envelope. These ``photon-bubbles'' are a phenomenon that
occurs in super-Eddington atmospheres and can support large density
inhomogeneities.

This reduces the continuum driving in the dense layers where clumps are optically
thick, thus allowing the stellar base to be hydrostatically bound even
though the luminosity is formally above the Eddington limit.
But as the clumps become optically thin in the higher-level, lower-density
layers, they are exposed to the full radiative driving of the assumed 
super-Eddington luminosity.
The intermediate layer where the net outward force equals gravity then becomes the 
sonic point of a wind outflow, with the density at this point setting 
the wind mass loss rate.
This identification of the sonic radius can be used 
to predict mass loss rates that are consistent with the observations of 
clear-cut super-Eddington objects \citep{s01}.  

For a given Eddington parameter $\Gamma$, the mass loss initiated in such
models depends on the nature of the assumed porosity.
Specifically, within the ``power-law porosity'' formalism 
developed by  Owocki, Gayley \& Shaviv (2004; hereafter \cite{ogs04}),
it depends on a power-index and the ratio of a
characteristic ``porosity length'' to the local scale height
(see \cite{ogs04} and section~2.4 below).
Analytic solutions derived by \cite{ogs04} show that physically
plausible porosity parameters combined with moderate super-Eddington
parameters, 
$\Gamma \gtrsim 1$ can indeed lead to mass loss rates comparable 
to that inferred for the extreme LBV giant eruption of $\eta$~Carinae,
even approaching the ``photon-tiring'' limit. 

A recent companion paper (van Marle, Owocki \& Shaviv 2008; hereafter
\cite{mos08}) used numerical radiation hydrodynamics simulations 
to test these analytic, porosity-moderated wind solutions of \cite{ogs04} 
for cases that do not exceed this photon-tiring limit.
The results showed that the asymptotic steady-states of the numerical
simulations are generally in quite close agreement with the analytic results.
However, with only modest variations in the choice of the key porosity
parameters, these analytic scaling laws also allow base mass fluxes
that can {\em exceed} the photon-tiring limit, with the consequence
that the initial wind solution must necessarilly stagnate at some
finite radius \citep{mos08a},  thereby precluding a net, steady mass outflow.

The central thrust of the current paper is to explore the
nature of the necessarily time-dependent flow that ensues from 
such a porosity model with a base mass flux above this tiring limit.
In principle this should involve complex, three-dimensional
(3-D) patterns of both outflow and infall. 
But given the computational and conceptual challenges in developing
dynamically realistic 3-D description, 
we first explore here much simpler, albeit idealized one-dimensional
(1-D) models.
A key rationale for this approach is that the underlying phenomenon of 
photon-tiring is not inherently multi-dimensional, but rather depends largely
on the {\em radial} variation of driving vs. gravity. 
The insight gained in studying 1-D simulations of the complex,
time-dependent combination of outflowing and inflowing layers will
form a basis for future, more complete models that include
multi-dimensional (2-D or even 3-D) flow.

As detailed in \S 2 (and also the appendix), a key feature of our
simplified method here is to maintain a global energy conservation that
accounts both for the loss of radiative luminosity in regions of
outward acceleration, as well as the re-energization of the luminosity
from regions of infall and shock heating.
The results in \S 3 illustrate the complex pattern of inflow and
outflow, including also the strong effect on the emergent radiative
luminosity. 
We follow in \S 4 with a discussion of what these results mean for
understanding the nature of LBVs, and then conclude in \S 5 
with a summary of the work thus far and a brief discussion of
directions for future work.
 
\section{Method}
\label{sec-annum}
\subsection{Mass and momentum conservation equations}

The spherical mass loss simulations presented here are based on evolving the 
1-D, time-dependent equations for conservation of mass and momentum.
For density $\rho$ and radial flow speed $v$, mass conservation
in radius $r$ and  time $t$ 
requires
\begin{equation}
\frac{D \rho}{D t} \equiv
 \frac{\partial \rho}{\partial t} +
 v \frac{\partial \rho}{\partial r} =
 - \frac{\rho}{r^{2}} \, \frac{\partial ( v r^{2 })}{\partial r}
 \, ,
\label{masscon}
\end{equation}
where $D/Dt \equiv \partial/\partial t + v \, \partial/\partial r$ 
is the total time derivative following the flow.
The total change in speed  depends in turn on the net force-per-unit
mass from gas pressure $P$, gravity, and radiation
\begin{equation}
\frac{Dv}{Dt} \equiv
\frac{\partial {v}}{\partial t} + v \frac{\partial v}{\partial r} =
- \frac{1}{\rho} \, \frac{\partial P}{\partial r}
- \frac{ GM}{r^{2}}
+ g_{\mathrm{rad}}
,
\label{eom}
\end{equation}
where, following standard notation,
$G$ and $M$ are the gravitation constant and stellar mass,
and $g_{\mathrm{rad}}$ is the radiative acceleration, 
described further in \S 2.4.

The pressure obeys the ideal gas equation of state,
\beq
P = \rho \frac{kT}{\mu} \equiv \rho a^{2}
\label{pgas}
\, ,
\eeq
with the latter equality defining the isothermal sound speed $a$
in terms  of the temperature $T$, molecular weight $\mu$, and Boltzmann's
constant $k$.
Because $a \ll v_{\mathrm{esc}}$, where $v_{\mathrm{esc}} \equiv \sqrt{2GM/R}$ 
is the escape speed from the stellar surface radius $R$, 
gas pressure generally plays little direct role in the outward driving 
of a stellar wind mass loss.
But it does provide a basis for matching the outflow onto a
hydrostatically stratified atmosphere at the wind base.
Moreover, as discussed below (\S 2.2), when an outflow stalls or stagnates, 
gas pressure plays a key role in dissipating and stopping the 
subsequent gravitational infall.

As such, our simulations here retain a non-zero gas pressure, 
but to avoid the necessity of modeling in detail the evolution of the 
(relatively insignificant) gas internal energy, 
we simply assume the gas temperature is a fixed constant.
This is set to give a sound speed $a=20$~km/s,
implying a temperature roughly equal to the stellar effective temperature, 
$T \approx T_{\mathrm{eff}} \approx 50$,$000$~K.

\subsection{Photon tiring formulation}

A key feature of the present simulations is to account
properly for the {\em work} done by radiation in accelerating material
outward against gravity.
As detailed in the Appendix,
a full treatment should in general include the
associated time-dependent equations 
for conservation of {\em radiative} momentum and energy
[eqns. (\ref{radmom}) and (\ref{raden})],
as well as a description of the 
energy exchange between gas and radiation [eqn. (\ref{gasen})].
However, the further analysis there shows that,
by assuming a limit of infinite light speed in such a way
that both the propagation and diffusion times of radiation become
small compared to any competing dynamical scale,
the storage of energy in the radiation energy density $E$ can be
effectively bypassed in favor of direct, essentially instantaneous changes
in the radiative flux $F$.

This makes it possible to account for the net work done by the
radiation in terms of a simple ordinary differential equation (ODE) 
for the radial change of the radiative luminosity $L=4 \pi r^{2} F$
[eqn.\ (\ref{radeniso1d})],
\beq
\frac{dL}{dr} =  
- {\dot M} g_{\mathrm{rad}}
- 4 \pi r^{2} {\dot Q} 
\, ,
\label{dldrq}
\eeq
where the local mass flux, 
${\dot M} \equiv 4\pi \rho v r^{2}$, 
is not necessarily constant (or even positive) in a time-dependent flow.
%
The first term on the right side represents the change in luminosity associated
with the net work done by the radiative acceleration, while
the second term represents the luminosity change
associated with the volume heat exchange rate ${\dot Q}$ between
radiation  and the gas energy [see eqn.~(\ref{gaseniso1d})].

In regions of outflow, the local mass flux is 
positive, and so the first term represents the basic
``tiring'' effect that leads to a radial decline in the radiative
luminosity $L$.
Likewise, in such regions of outflow, the need to keep the gas temperature 
(and internal energy $e$) constant against the expansion cooling
requires ${\dot Q} >O$ [see eqn.~(\ref{gaseniso1d})], 
implying that the second term also contributes to a (relatively small) 
further radial reduction in luminosity.

In a time-dependent flow, however, the mass flux is not generally constant.
Indeed, in regions of flow stagnation and inflow, where $v<0$, it can even be 
locally {\em negative},  and therefore lead to a radial increase, or  
{\em re-energization}, 
of the luminosity in the regions above a local infall.
Moreover, such downflows are typically stopped eventually by strong, 
localized shock compressions, which
effectively convert the infall kinetic energy back into radiation,
making ${\dot Q} < 0$. 
This represents yet another mechanism for re-energization of the
luminosity, now occuring at radii above a shock compression.

Using 
eqn.~(\ref{gaseniso1d}) for the heating rate needed to keep the 
gas isothermal, the luminosity change can be alternatively written as
\beq
\frac{dL}{dr} 
=
- 4 \pi \rho v r^{2} g_{\mathrm{rad}} 
- 4 \pi P \frac{d(vr^{2})}{dr}
\label{dldr}
\, .
\eeq
%
This equation provides a convenient form to compute the
radial change of luminosity in a spherically symmetric, isothermal flow,
applicable even in the fully time-dependent cases considered below.
At each time-step of the simulation, this ODE can be
integrated numerically from the base radius $R$ to give the radial
run of luminosity, $L(r)$, which is then used to compute the
radiative acceleration $g_{\mathrm{rad}}$ for the next step of the simulation.
To avoid the unphysical possibility of a negative
luminosity, the integration of eqn.~(\ref{dldr}) sets a floor luminosity $L=0$.

In practice, maintaining a global energy conservation when there are
regions and intervals of infall requires one to account also for the
energy associated with material that falls back onto the stellar
surface through the lower boundary radius $R$. 
We minimize this here by limiting the inflow velocity at the boundary to
the sound speed, which then generally leads to shock formation and
recapture of most of the infall kinetic energy through the instantaneous 
emission terms described above.
Nonetheless, there is still a small energy loss through the
boundary that has to be accounted for to maintain global energy
conservation.
We do this here by accumulating a storage energy 
$E_{ac}$ that is then assumed to be re-emitted on a 
parameterized time scale $t_{em}$, taken here to be several
(viz. 10) wind flow times.
Based on the value of the stored energy at any given time step,
the lower boundary condition for the luminosity integration of
eqn.~(\ref{dldr}) then becomes
\beq 
L(R) = L_{\ast} + \frac{E_{ac}}{t_{em}}
\, .
\label{llbc}
\eeq
Generally, the associated modification in base luminosity is small,
no more than a few percent.

\subsection{Tiring in steady-state outflows}

For a simple, steady-state flow with constant mass flux ${\dot M}$, 
we can rewrite the radial change of luminosity in the forms,
\beqa
\frac{dL}{dr} 
&=& - {\dot M} 
\left [ g_{\mathrm{rad}} + P \, \frac{d}{dr}\biggl(\frac{1}{\rho}\biggr) \right ]
\\
&=&  - {\dot M} 
\left ( \frac{GM}{r^{2}} + v \, \frac{dv}{dr} \right )
\, ,
\label{dldrss}
\eeqa
where the latter equality assumes an isothermal flow and uses the
momentum equation (\ref{eom}) to eliminate the radiative acceleration,
with the result that the net tiring no longer depends explicitly on
the pressure work term.
Upon integration from the base radius $R$, where flow speed is
negligibly small ($v(R)  \ll a$) and the luminosity is set to the 
interior value, $L(R) = L_{\ast}$,
the radial dependence of luminosity is given by
\beq
L(r) = L_{\ast} - {\dot M} 
\left [ \frac{v(r)^{2}}{2} + 
\frac{GM}{R} \left ( 1 - \frac{R}{r} \right )  
\right ]
\, .
\label{lss}
\eeq
Note in particular that, in this approximation of isothermal flow with
constant internal gas energy, the pressure work term makes {\em no} net
contribution to the change in luminosity.

The form (\ref{lss}) is in fact the form used in the
steady-state, continuum-driven models derived in \cite{ogs04}.
It also helps justify neglecting the pressure work terms in the
time-dependent simulations of paper 1, which were used to relax towards the
asymptotic steady-states.

For a wind with terminal speed $v_{\infty}$, the emergent luminosity is
\beq
L_{\infty} = L_{\ast} - {\dot M}
\left [ 
\frac{v_{\infty}^{2}}{2}
 + \frac{GM}{R} 
\right ]
\, .
\label{linf}
\eeq
Setting $L_{\infty} = v_{\infty} = 0$ then defines a
photon-tiring-limit for the maximum possible
mass loss rate for a steady-state outflow
\citep{og97, ogs04},
\beq
{\dot M}_{\mathrm{tir}} \equiv \frac{L_{\ast}}{GM/R}
\, .
\label{mdtir}
\eeq
Any flow with base mass flux greater than this  limit must necessarily
stagnate at a finite radius.
The simulations presented below examine the complex, time-dependent
mixture between outflow and infall that results in flows above
this tiring limit.

\subsection{Porosity-moderated continuum driving}

The treatment of radiative driving in the time-dependent
simuations here is based on the formalism for 
porosity-moderated, continuum driving developed for the 
steady-state analysis of \cite{ogs04}.
For a local luminosity $L(r)$, the radiative acceleration is given by
\beq
g_{\mathrm{rad}} (r) = \frac{\kappa_{\mathrm{eff}} L(r)}{4 \pi r^{2} c}
\, ,
\eeq
where $c$ is the speed of light, and $\kappa_{\mathrm{eff}}$ is an {\em effective}
opacity that is reduced from the microscopic value $\kappa$
by the porous nature of the medium.
For a power-law distribution of clump strengths
characterized by a power index $\alpha$ and a ``porosity length'' 
$h_{o}$ that sets a maximum clump optical thickness 
$\tau_{o} = \kappa \rho h_{o} \equiv \rho/\rho_{o}$,
the opacity reduction factor depends on the density through
[cf. \cite{ogs04} eqn.~(57)],
\beq
k (\rho) \equiv \frac{\kappa_{\mathrm{eff}}}{\kappa} =
\frac{ (1+\rho/\rho_o)^{1-\alpha} - 1}{ (1-\alpha) \rho/\rho_o } 
\, .
\label{kdef}
\eeq
Note that for low densities $\rho \ll \rho_{o}$,
the clumps all become optically thin, making $k \rightarrow 1$, 
since the material is exposed to the full opacity $\kappa$.
However, at very high densities $\rho \gg \rho_{o}$, self-shadowing of
optically thick clumps reduces the effective opacity, following an overall
power-law dependence on density, 
$k \sim (\rho_{o}/\rho)^{\alpha}$.
 The self-shadowing of optically thick clumps is inherently a 
multi-dimensional effect at the microscopic level, 
but at the microscopic level appears as 1D.

Even in a star for which the continuum opacity $\kappa$ implies an
Eddington parameter $\Gamma \equiv \kappa L/4 \pi GMc$ above the 
Eddington limit, i.e. $\Gamma > 1$, 
the porosity reduction of the radiative force can allow 
high-density regions to remain gravitationally bound, while the outer,
lower-density regions are driven into a wind outflow.
The transition occurs at the sonic point, where $\Gamma k[\rho_{s}] = 
1$, thus defining a base mass loss rate 
${\dot M} = 4 \pi \rho_{s} a R^{2}$.
For the canonical case $\alpha=1/2$ assumed in the simulations here, 
this has the scaling [cf. \cite{ogs04} eqn.~(77)],
\beq
{\dot M} = 4 (\Gamma - 1) \, \frac{L_{\ast}}{\eta \, ac} 
\, ,
\label{mdpow}
\eeq
where $\eta \equiv h_{o}/H$, 
with $H = a^{2} R^{2}/GM$ being the density scale height at the subsonic 
base of the wind.
The ratio of this mass loss rate to the tiring 
limit given in eqn.~(\ref{mdtir}) 
has the scaling [cf. \cite{ogs04} eqn. (78)],
\beq
m_{\mathrm{tir}} \equiv
\frac {\dot M}{\dot M_{\mathrm{tir}}} = 0.13 \, 
\frac{\Gamma - 1}{\eta \, a_{20}} \, 
\frac {M/M_{\odot}}{R/R_{\odot}}
\, ,
\label{mtirnum}
\eeq
where $a_{20} \equiv a/20~$km/s.

\subsection{Parameters and numerical implementation}
\label{sec-input}
The simulations here are all based on a single
representative stellar model with radius $R=50 R_{\odot}$,
and mass $M=50 R_{\odot}$, implying a solar value for the escape speed,
$\vesc=617\kms$. 
As noted, the sound speed is fixed at $a=20\kms$, corresponding
roughly to a temperature of ca.~50,000~K.

A summary of the varied model parameters is  given in Table~1.
Models A-C assume a stellar luminosity  
$L_{\ast}=1.6 \times 10^{7} L_{\odot}$, 
which with an assumed opacity $\kappa=0.4 \, {\rm cm^{2}/g}$, 
implies a strongly super-Eddington condition with $\Gamma = 10$.
Model D assume a moderate $\Gamma=5$,
while Models E \& F assume a marginally super-Eddington condition 
$\Gamma = 2$.
The models are further distinguished by the choices of the porosity
length parameter $\eta$, as specified in third column of Table~1.
For the assumed porosity power index $\alpha_{p}=1/2$,
eqn.\ (\ref{mtirnum}) shows that the values of $\eta$ and $\Gamma$ combine 
to yield a predicted mass-loss tiring ratio $m_{tir}$, 
as given in the fourth column of Table~1.
As shown below, the level by which this base mass-loss tiring exceeds 
unity is key to the overall nature of the flow stagnation and
infall of the various models.



For all cases, we set the re-release timescale for the inner-surface
infall energy
to be 10 stellar-freefall times,
$t_{\rm em} = R/\vesc \simeq 5.5\times10^5$s.
Since the energy storage and relative luminosity contribution is
generally small compared to the stellar luminosity, we find the
results are generally insensitive to the specific release time chosen.


As in Paper~1, the simulations here 
use the ZEUS~3D code developed by \citet{sn92} and \citet{c96}. 
But unlike the purely isothermal models previously relaxed to an asymptotic
steady-state, the complex time-dependent 
evolution of photon-tired flows here requires keeping careful 
account of the energy exchange, $\dot{Q}$, between the gas and
radiation. 
Our implementation of ZEUS achieves this by advancing each individual
time step assuming adiabatic conditions, with an adiabatic index 
$\gamma = 5/3$, but then resetting 
the temperature at each grid point back to the assumed constant value.
The difference in internal energy between the constant temperature and
the temperature that follows from the adiabatic calculations is then
added to or subtracted from the local radiative luminosity, 
following eqn. (\ref{dldrq}).

In wind simulations, a key general issue is to resolve the acceleration
from a nearly hydrostatic layer supported by gas pressure, including
for example, at the subsonic wind base, which has a characteristic 
scale height $H \approx (2 a^{2}/\vesc^{2}) R = 0.002 R$.
However, in the models here, the repeated flow stagnation can often
lead to multiple such static layers distributed over many stellar
radii from the wind base.
Moreover, the ultimate escape from the system is only ensured once the
outward flow speed exceeds the local escape speed, which can sometimes
extend to over 100~$R$.
To properly resolve this complex flow over a large spatial range, we
use $n_r = 20,000$ spatial zones extending over 
$r=1-200$~R.
To concentrate resolution more in the inner wind, we use a spatial
grid size that increases by a factor 1.0005 per zone, i.e. 
$dr_i = dr_{i-1}*1.0005$.
This leads to a spatial resolution that ranges from 
$dr_{0} \approx 5 \times 10^{-6} \, R$ at $r=1 \, R$ to 
$dr_{nr} = 0.1 \, R$ at $r=200\, R$.

Finally, as in Paper~I, the simulations start with an initial condition set
by a simple $\beta$-law velocity profile, 
\beq
\nonumber
  v(r)=v_\infty \left ( 1-\frac{r}{R} \right )^\beta,
\eeq  
with $\beta=0.5$,
and a mass flux defined by the porosity length formalism [eqn.~(\ref{mdpow})]. 

The boundary conditions are the same as used in \cite{mos08}.
The density at the inner boundary is fixed to a value about 10 times
the sonic point density for the analytic porosity model.
The velocity itself is allowed to float, requiring only that the
velocity gradient remain constant across the inner boundary, and that the
absolute value of the base velocity is limited to the sound speed. 
Such boundary conditions have been used succesfully in simulations of 
line-driven  winds simulations in both one \citep{ocr88} and 
two dimensions \citep{ocb94}.

\begin{figure*}
\begin{center}
\resizebox{\hsize}{!}{\includegraphics[width=\textwidth,angle=-90]{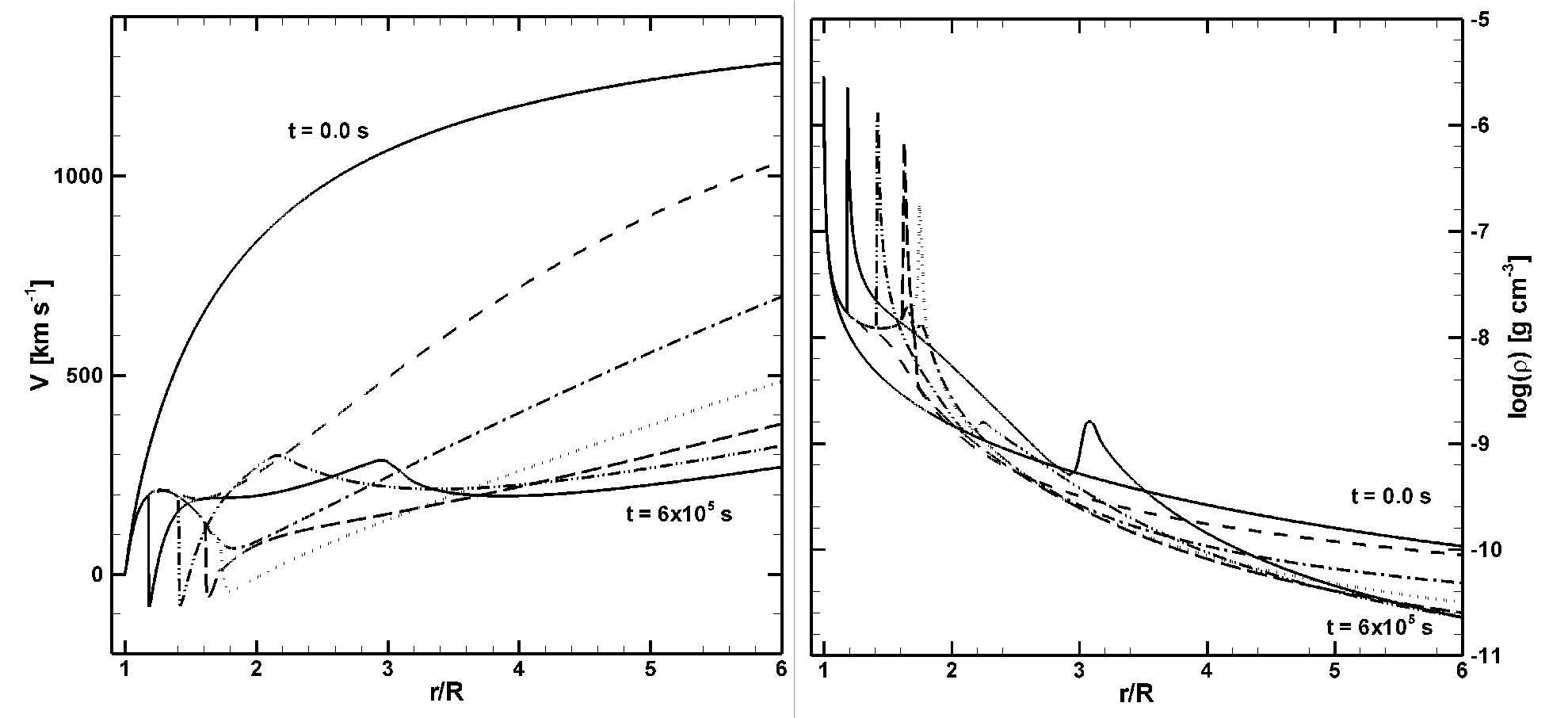}}
\caption{
Snapshots of the velocity (a) and log density (b) vs. radius at multiples of 
$10^{5}$s after the switching on of photon tiring, at $t=0$, for Model B. 
The model was first relaxed to a steady state ignoring tiring.
The reduction in the radiative driving flux induces a slow-down towards flow
stagnation, with shock collisions leading to a build up of high-density shells 
that eventually fall back onto the star.
}
\label{fig:vddrop}
\end{center}
\end{figure*}    

\section{Results}
\label{sec-result}
  
\subsection{Initial relaxation and emergence of photon-tiring}

As a first illustration of the effect of photon tiring, figs.
\ref{fig:vddrop} plot the radial variation of velocity (a) and 
log-density (b) at fixed time snapshots after turning on the tiring
effect in a model which was first relaxed to a steady-state without tiring.
The model parameters are those of model~B, with a base mass flux 
 $m=2.34$ times the tiring limit.
Shortly after the tiring effect is implemented, the existing 
flow begins to decelerate, leading to stagnation and then infall back 
toward the star. 
As this infalling material meets the remaining upflow below,
strong shocks form, compressing the material into dense shells, 
marked by the density spikes in fig. \ref{fig:vddrop}b.
Since these dense shells have too much mass to be driven outward by 
the available radiative flux, they are pulled back by gravity, and
re-accreted back onto the star.

Eventually, some of the energy generated can lead to a net outward
driving of some of the gas, but it too can stagnate before being able to escape, 
forming more shock layers, and so forth.
The actual details are of course complex and depend on the specifics of the 
overloaded initial condition.

However, rather than dwelling on the specific details which depend on the initial
readjustment, we shall proceed to focus on the steady-state
flow properties, long after the wind relaxes from its initial conditions.

The timescale over which this relaxed state is reached is on the order 
of several times $10^8$ seconds. 
This is effectively the crossing time for gas elements with
a characteristic velocity of approx.~$50\,\kms$ (the flow speed at the
outer boundary) over the length of the grid (200 stellar radii).

\begin{figure*}
\begin{center}
\resizebox{\hsize}{!}{\includegraphics[width=\textwidth,angle=-90]{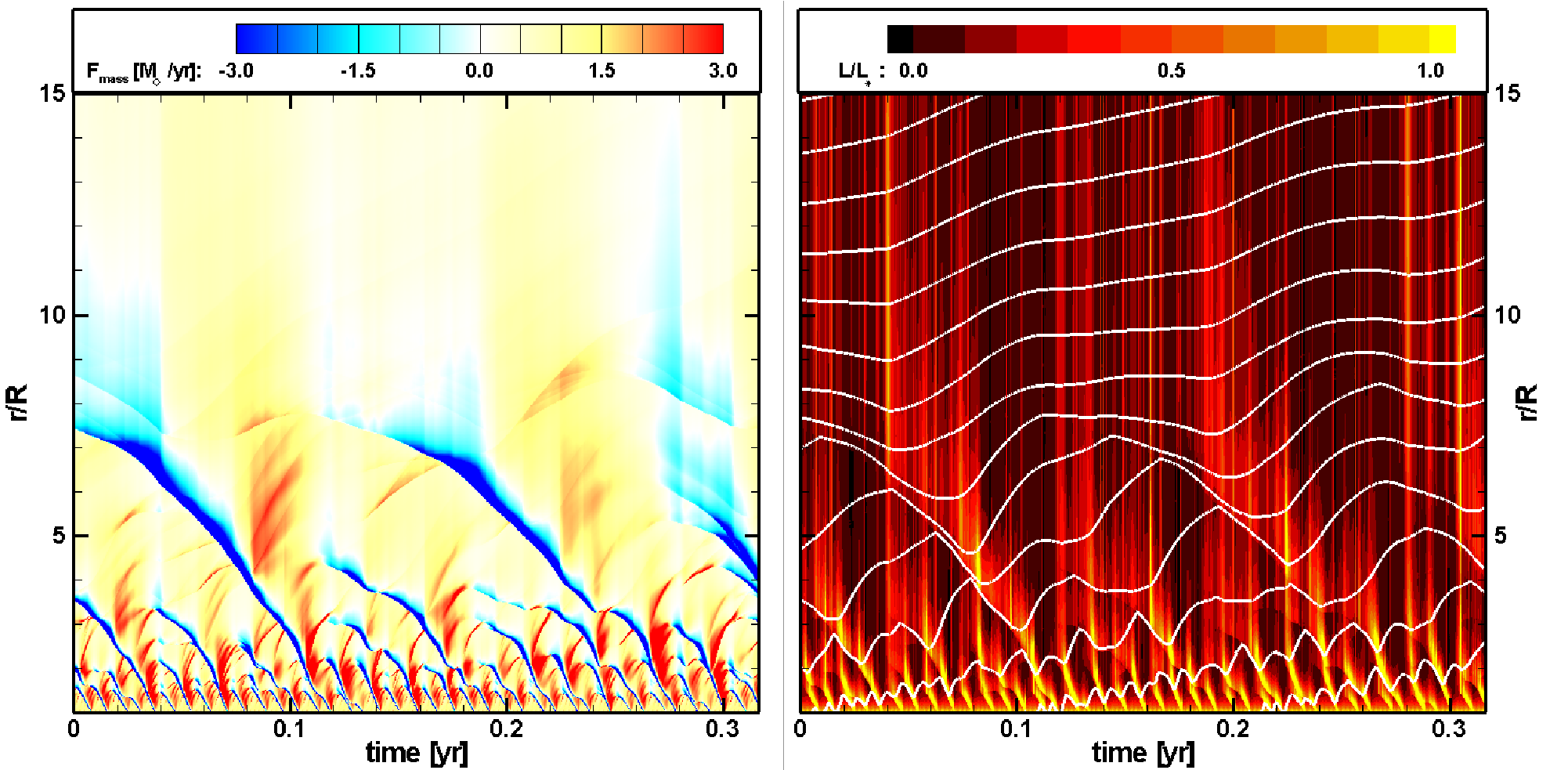}}      
\caption{
Left: 
Colorscale showing mass flux vs.\ radius and time for Model B. 
Mass leaves the surface layers only to fall back. 
Part of the mass reaches a higher altitude, where the process repeats
itself several times, until the gas can coast away from the star and
eventually reach the local escape speed.
%
Right: Colorscale showing the associated 
radiative luminosity as a function of time and
radius.  
The superposed lines depict the position of
individual mass elements (constant Lagrangian mass).
The outward motion of the mass elements increases whenever the luminosity is high 
and slows down, or even reverses if the luminosity is low.
But eventually, mass parcels at large radii move systematically 
outward.
}
\label{fig:flmass}
\end{center}
\end{figure*}

\subsection{Steady-state behaviour of the  shock layers} 

To illustrate the complex pattern of infall and outflow in the regions
with the overloaded wind,
fig.~\ref{fig:flmass}a presents a colorscale plot of the mass flux for model~B 
(base mass flux $2.34\times\dot{M}_{\mathrm{tir}}$) over the radial range $r = 1-15$~R 
and in a representative time interval of 
about a third of a year, long after relaxation from the initial conditions is reached.
The figure shows quite vividly the complex hierarchy of initial base outflows
(in red) that are truncated by broad downflows (in blue) of stagnated material.
This flow pattern is a direct consequence of the photon tiring feedback.

Indeed, the colorscale in fig.~\ref{fig:flmass}b reveals that this flow pattern is
tightly coupled with corresponding time and space variations of the
local luminosity.
The photon tiring in regions of outward acceleration leads to a strong reduction
in the luminosity (dark shading);
but this is punctuated by re-energization of this luminosity (light
shading) in the shock collision regions where infall energy is converted back
into radiation.

To further demonstrate the overall spatial progression of mass 
in this complex flow pattern, the overlaid light contours in
fig.~\ref{fig:flmass}b mark the radial and temporal evolution of 
selected mass shells.
Because there can be no crossing of such mass shells in these
idealized, 1-D, purely radial flow simulations, they can be
effectively tracked by defining the time and radius variation of a
Lagrangian mass coordinate,
\begin{eqnarray}
m_{\rm L}(r,t) & & \equiv~4 \pi \int_{r}^{R_\infty}  r'^2\rho(r',t) dr' \nonumber \\
	   & &+~4\pi \int_0^t  r^2 \rho(r,t') v(r,t') dt' ,
\end{eqnarray}
with $R_\infty$ being the outer boundary of the grid \citep{ocr88,op99}.
The first integral labels each mass shell of the initial ($t=0$)
condition by the amount of mass to the outer boundary;
the second integral then tracks the amount of mass that has passed through a zone
with radius $r$ in the time $t$ since that initial condition.
The mass shell tracks plotted in fig.~\ref{fig:flmass}b therefore
represent contours of this Lagrangian mass vs. radius and time.

Together, the plots in fig.~\ref{fig:flmass} provide an interesting
complementary perspective for the flow patterns in models with a 
base mass flux above the tiring limit.
Matter is launched from the surface of the star, only to fall back
almost immediately as the radiation field becomes overloaded and can
no longer support the outflow against gravity.  
Once, however, the infalling gas is shocked, it releases
its kinetic energy as radiation, which can then provide a boost to the 
overlying material. This implies that some of the gas can accelerate above the first 
shock layer.

Nevertheless, because the gas accelerating above the first layer is too overloaded, 
it will again 
stagnate and begin to fall back. The energy it releases can then accelerate gas above 
the second shock, and so forth. 
However, because progressively higher shocks are further from the star and involve a 
smaller mass flux, their time scale for acceleration and stagnation is longer.

At a high enough layer, both the total mass flux and the gravitational potential depth 
are small enough such that the remaining luminosity can drive the gas to infinity. 
Beyond this layer, a freely coasting wind is accelerated to infinity. 
In model B, for example, this takes place at roughly 8 stellar radii.  
 
\begin{figure}
\resizebox{\hsize}{!}{\includegraphics[width=\columnwidth,angle=0]{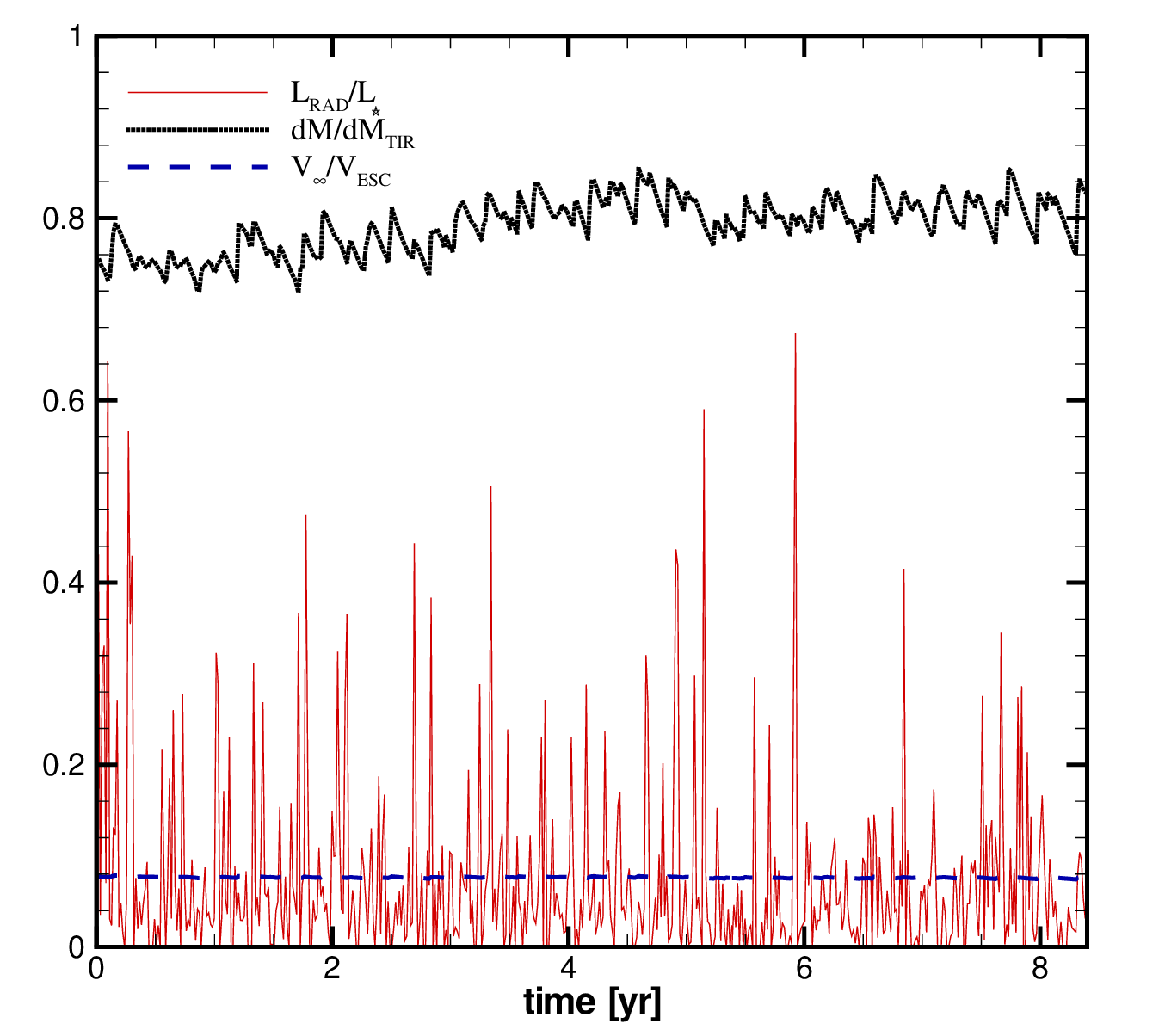}}      
\caption{
Mass flux relative to the tiring mass loss rate (jagged black line), wind
velocity relative to the escape speed at the stellar surface (dashed
line) and radiative luminosity relative to the stellar luminosity
(continuous line) at the outer boundary radius ${200\, \Rstar}$ 
for model B with ${\dot{M}=2.34\dot{M}_{\rm tir}}$, 
after the simulation has fully relaxed.  
The mass flux varies slightly below to the tiring limit, while the wind
velocity is nearly constant at a value well below the surface escape
speed.
The emergent radiative luminosity is highly variable,
but with an average value that is only about 10\% of the base
luminosity, i.e., $\Gamma_{\infty} \sim 1$.
}
\label{fig:out050}
\end{figure}    
  
\subsection{Wind behaviour at larger radii} 

To demonstrate the nature of this net outflow at larger
radii, fig.~\ref{fig:out050}  plots the time variation of three key
quantities at the outer boundary  of the computational grid
(${200\, \Rstar}$) for the relaxed state of model~B.
These include the radiative luminosity relative to the stellar luminosity, 
the wind velocity relative to the escape speed, and 
the mass flux relative to the photon-tiring mass loss rate.
The driving proves to be quite efficient, bringing the mass flux close to
($\sim85\%$) the tiring limit.
The wind velocity is nearly constant at ca. 50$\kms$, about a factor 2.5 times the
sound speed, and much less than the surface escape speed, $v_{\mathrm{esc}} (R) 
\approx 620$~km/s;
but it is just slightly above the local escape speed, $v_{\mathrm{esc}}
(r=200\, R) \approx 44$~km/s.
The mass flux is also quite constant, with the small fluctuations
that mostly reflect residual variations in the flow density from the
discrete shells formed near the wind base.

However, note that, at least in this idealized 1D model, the
radiative luminosity exhibits quite strong variations.
The average level is also subtantially reduced, 
to about $\sim 10\%$ of the base luminosity.
But there are also brief spikes above 50\%, arising from the sudden
radiative emissions in the shock infall regions.
More realistically, such spikes would likely to be smoothed out by
both radiative storage and 3-D averaging effects ignored in the present
1-D, sudden radiative release simulations.

\begin{figure*}
\resizebox{\hsize}{!}{\includegraphics[width=\textwidth,angle=0]{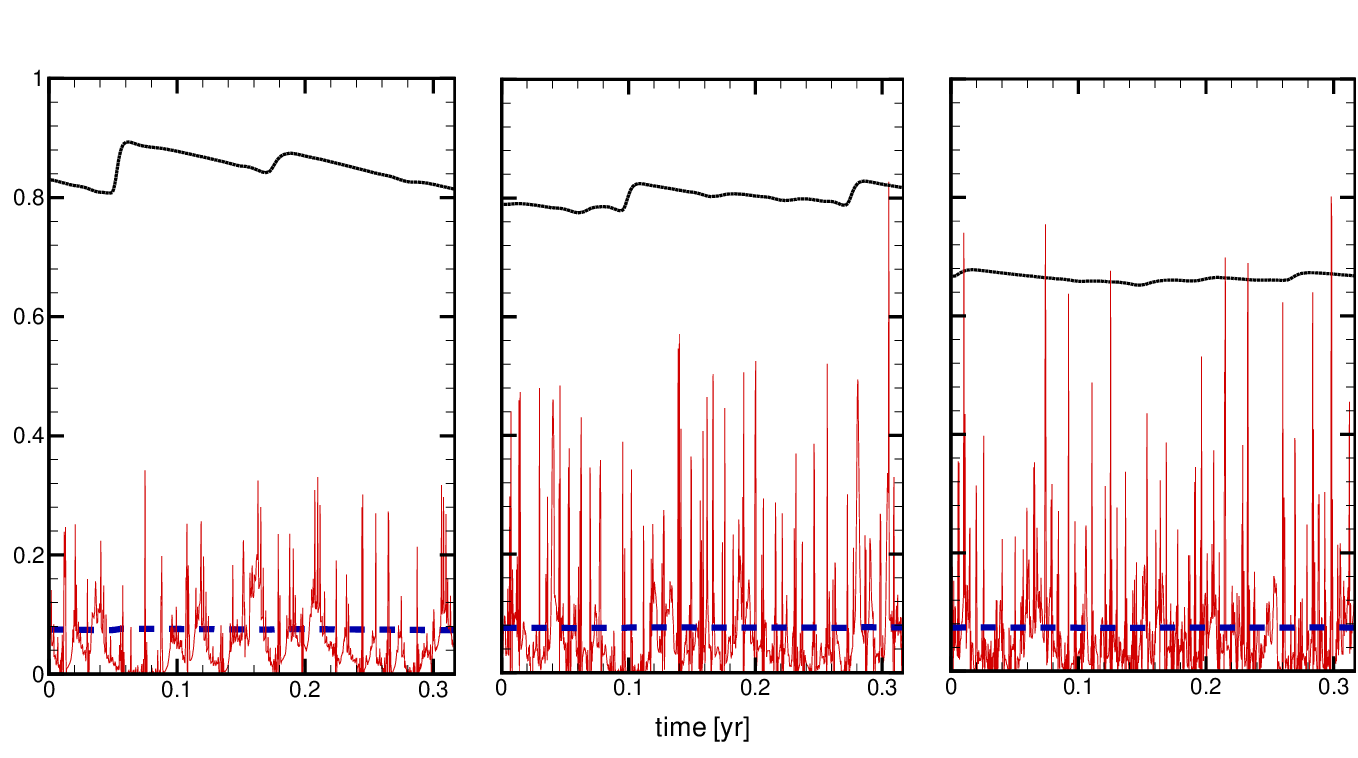}}      
\caption{
The same variables as fig.~\ref{fig:out050} for models C, B, and A, having (from
left to right) $\dot{M}$=1.17, 2.34 and $4.68\dot{M}_{\rm tir}$
respectively.
As the mass flux at the surface exceeds the tiring limit by a larger margin, 
the tiring mechanism becomes somewhat less efficient:
the mass flux at the outer boundary (jagged black line) decreases and radiative 
luminosity (spiky, continuous line) increases.  
Also, the variations in mass flux smooth out and decrease in size.  
The wind speed (dashed line) remains nearly the same for all
three simulations.}
\label{fig:triplot}
\end{figure*}

Finally, fig.~\ref{fig:triplot} compares the outer boundary state of 
models C, B, and A with increasing mass flux tiring factors, 
1.17, 2.34, and 4.68.
To show finer details of the variations, the time
resolution in fig.~\ref{fig:triplot} is much higher than
in fig.~\ref{fig:out050}.
Qualitatively, the behaviour of the wind is similar for all three
simulations: 
the mass flux is a large fraction ($>65\%$ of the tiring limit in all three
cases).
The wind moves slowly, and and the radiative luminosity that escapes at the
outer boundary is highly variable  but greatly reduced from the base value.

The results also reveal notable trends in the quantitative
properties.
In particular, for models that exceed the tiring limit by a greater margin,
the net driving of mass loss becomes somewhat less efficient.
More photons escape while the mass flux decreases and becomes less variable.
This is understandable as the motion of matter back and forth close to
the star recycles energy from the radiation field, through kinetic and potential
energy of the wind, and back into radiation. 
If the wind exceeds the tiring limit by a wider margin, this
process become more extensive, which reduces the efficiency of the driving
and allows more photons to escape. 
However, the outer wind velocity is almost the same in all three
models.
  
\begin{figure}
\resizebox{\hsize}{!}{\includegraphics[width=\textwidth,angle=0]{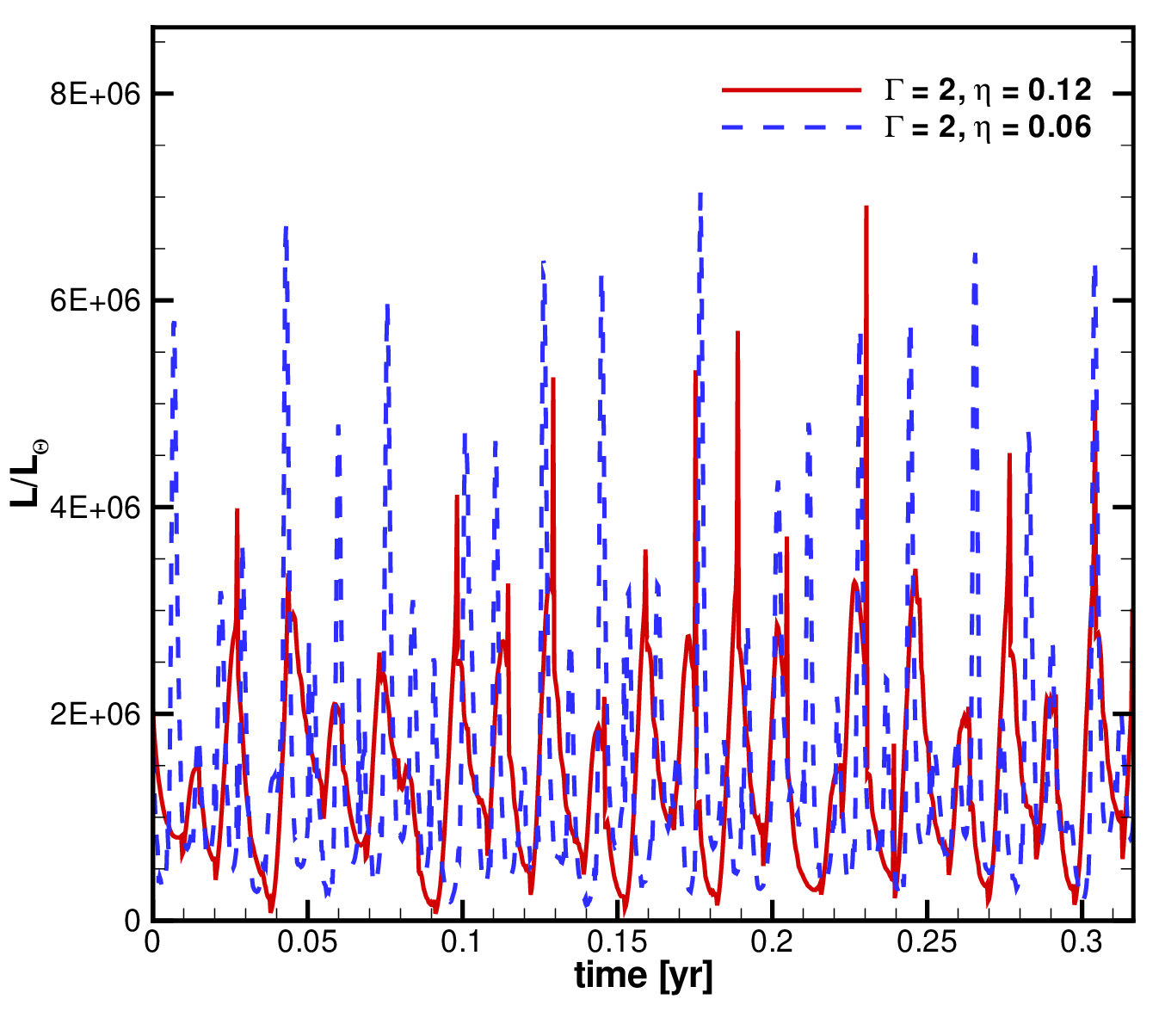}}      
\caption{
The light curve of two slightly photon-tired cases, with $\eta=0.06$ and $\eta=0.12$. 
Both have $\Gamma=2$. 
Unlike the higher $\Gamma$ models, the light-curves are much more regular, and not chaotic.
}
\label{fig:lowGammaLight}
\end{figure}

\begin{figure}
\resizebox{\hsize}{!}{\includegraphics[width=\textwidth,angle=-90]{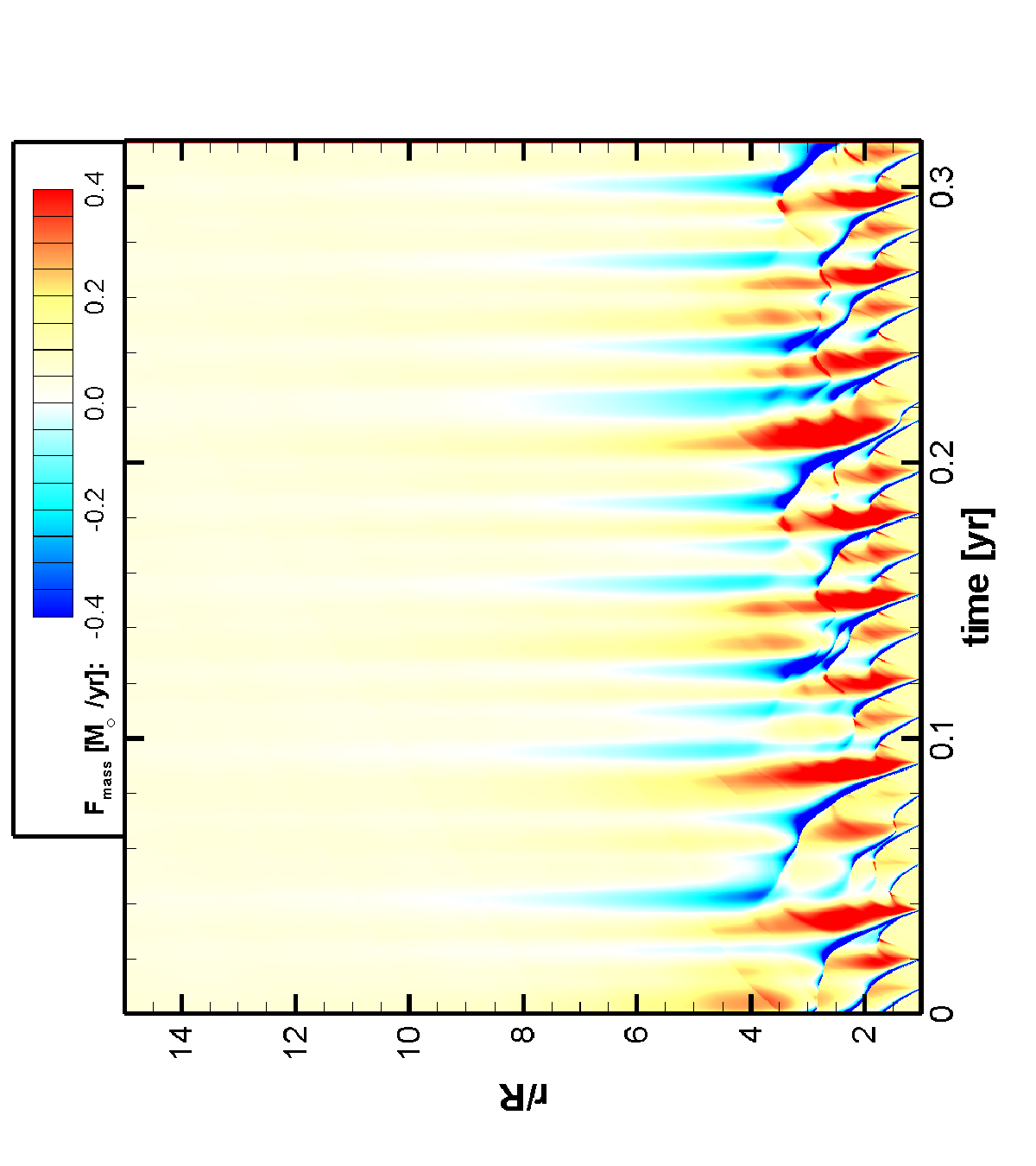}}
\caption{
The radial dependence of the mass flux as a function of time (i.e., similar to 
fig.\ref {fig:vddrop}b), 
for model F, with $\Gamma=2$ and $M/M_{tir}=1.08$. 
Because of the more moderate conditions, the shock layer structure is much simpler--it 
includes fewer layers, 
extends to a smaller radius, and significantly less chaotic 
(as can be seen in fig.\ \ref{fig:lowGammaLight})
}
\label{fig:lowGammaMassFlux}
\end{figure}

If we compare models with a high $\Gamma$ and high tiring number, 
to models with low $\Gamma$ and low tiring number, then we see a clear trend.
With a smaller tiring number, there are fewer shock layers up to where
free outflow is reached. 
As a consequence, the light curve is much more regular, 
as can be expected since it is composed of fewer shock ``oscillations". 
This can be seen in figs.\ \ref{fig:lowGammaMassFlux} and \ref{fig:lowGammaLight}.
A lower luminosity (lower $\Gamma$) means that a smaller amount of mass is present 
in the grid at any given time. 
Therefore, the optical depth is smaller, so the photons have a better chance to escape.
As a result, lower luminosities at the base generally lead to lower efficiency in 
the driving. 

In Table~1, we show the relative contributions of 
potential energy luminosity ($\dot{E}_g=1/2\dot{M}(\vesc(R)^2-\vesc(r)^2)$ and 
radiative (L) luminosity to  the total energy outflow ($\dot{E}$) at the outer boundary. 
These two make up the bulk of the energy flow, because kinetic energy flow 
($1/2\dot{M}v_\infty^2$) and the internal energy advection ($\dot{M}kT/m_H$) together
contribute less than 0.5\%.
Maintaining a fully accurate energy conservation in these simulations can be difficult due to 
the numerical problems that occur when resolving a large number of radiative shocks. 
We achieved an accuracy of more than 90\% in all simulations with the 
exception of simulation~A, which has the largest number of shocks, and
for which the energy conservation accuracy 
was about $\sim~75$\%. 
Therefore, although the results of the simulation seem to follow the trends observed 
in this section, the quantitative values should be viewed with some
caution.
  
\begin{table*}
\caption{Summary of model parameters and results. 
The other model parameters are $M = 50 M_\odot$, $R= 50 R_\odot$, 
$\kappa =0.4 cm^2~gr^{-1}$ giving $L_{\mathrm{Edd}} = 1.6~10^6 L_\odot$, 
$a= 20~km~s^{-1}$ and $\alpha_p = 1/2$. 
For the $\Gamma=10$ models, $M_{\mathrm{tir}} = 0.52~M_\odot/yr$. 
$n_{\mathrm{max}}$ is defined as the maximum number of radial shock layers, 
which quantifies the complexity of the radial structure, 
while $r_{top}$ is the largest radius to which shocks reach. $L/\dot{E}$ and 
$\dot{E}_g/\dot{E}$ are the contributions of radiative luminosity and gravitational 
potential energy respectively to the energy outflow at the outer boundary.}
\begin{tabular}{ccccccccccc}
\hline
Model & $\Gamma$ & $\eta$ & $\dot{M}_0/M_{\mathrm{tir}}$  & 
$\overline{\dot{M}_{\infty}}/M_{\mathrm{tir}}$ & 
$\overline{\Gamma_{\infty}}$ & $n_{max}$ & $v_{\infty}/v_{\mathrm{esc}}$ & 
$r_{top}/R$ & $L/\dot{E}$ & $\dot{E}_g/\dot{E}$ \\
\hline 
A &  10  &  0.25  &  4.68   &   0.66    &   0.7  &    8   &  0.07      &  6  & 0.094  & 0.90\\
B &   10  &  0.5  &  2.34   &   0.8     &   0.7  &    5   &  0.08      &  8  & 0.076  & 0.916\\
C &   10  &  1.0  &  1.17   &   0.85    &   0.5  &    4   &  0.08      &  8  & 0.059  & 0.935 \\
D &   5  &  0.25  &  2.08   &   0.68    &   1.5  &    5   &  0.08      &  5  & 0.307  & 0.689\\
E &   2  &  0.06  &  2.17   &   0.38    &   4.4  &    4   &  0.09      &  2.8  & 0.466  & 0.527\\
F &   2  &  0.12  &  1.08   &   0.47    &   4.3  &    3   &  0.09      &  3.5  & 0.450  & 0.544\\
\hline
\end{tabular}
\end{table*}
 
\section{Discussion}
\label{sec-concl}

It should be emphasized here that the photon-tiring effects that are
the focus of this paper are only relevant for the most extreme forms
of radiatively driven mass loss.
For even the strongest, steady-state, line-driven winds from WR stars,
the mass loss rates can range up to a few times 
$10^{-5} \, M_{\odot}$/yr and the flow speeds up to 
$v_{\infty} \approx 3000 \,$km/s.
This then implies a mechanical wind liminosity that is less than a few
percent of a typical WR luminosity of ca. $10^{6} \, L_{\odot}$.

By contrast, the Homunculus nebulae that resulted from 
the 1840-1850 giant eruption of $\eta$~Carinae is estimated to harbor
at least $10 \, M_{\odot}$ with expansion speeds on order $600 \,$km/s
\citep{s02}, implying a characteristic average mechanical luminosity that
is quite comparable to the estimated radiative luminosity 
$2.5 \times 10^{7} \, L_{\odot}$ during the epoch.
This seems a strong indication that photon tiring may have played a
key role in the driving of this extensive mass loss.

But there are some key differences between these properties of
the $\eta$~Carinae eruption and the 1-D photon-tired wind models
computed here.
For one, the  observed terminal flow speed of ca. 600~km/s 
is much higher than the ca. 50~km/s found for the tiring-limited
models computed here.
Moreover, in contrast to the rough parity between mechanical and
radiative luminosity for $\eta~$Carinae's giant eruption,
these tiring models give mechanical luminosities that are several
times higher than the emergent radiation, which itself is also quite
variable.

However, as already noted, it seems quite likely that these general
properties of 1-D photon-tiring could be quite different in full 3-D
case.
In particular, the likely lateral variations between regions of infall
and outflow could allow a stronger, more regular escape of radiative
flux. Indeed it might even resemble the extensive, complex 3-D 
clump structure envisioned in the parameterized porosity formalism.
The net effect could be a self-regulation of porosity to keep the mass
flux below the tiring limit, and thus allow the kind of
quasi-steady, higher-speed wind outlfow obtained in both the analtyic models
of \cite{ogs04} and the numerical simulations of \cite{mos08}.
  
The up-flow and down-flow nature of the mass flux close to the star is also 
reminiscent of convective motion, albeit driven by radiative luminosity 
rather than buoyancy.
In the stellar interior, convection can effectively transport energy
at such a rate that the star can locally exceed the Eddington limit
\citep{jso73}.
But in surface layers it becomes inefficient due to the lower 
density \citep{ogs04}.
Similarly, the shock layers allow for mechanical energy advection to keep the photon-tired 
layer with a reduced luminosity. 
However, unlike convection where the flows are sub-sonic, 
the photon tired layer is composed of highly super-sonic motion. 

Even in the 1-D framework, the present simulations assume an 
instantaneous nature for both the propagation and diffusion of
radiation, and neglect optically thick backwarming effects in favor of a 
simple, constant temperature.
Including an optically thick radiative diffusion with a finite speed
would allow energy to be stored in the radiation field, with a likely
smoothing effect on both luminosity variation and the resulting
motions of mass shells.
%
In 3-D or even 2-D such shells should break up into clumps. 
This could allow more photons to escape between the dense clumps,
reducing the mass flux and increasing the radiative luminosity at the
outer boundary.
This higher luminosity at large radii could also accelerate the flow to 
higher terminal speeds.

\section{Summary \& future work}
\label{sec-sum}
If the base mass flux from a star exceeds the photon-tiring limit, 
its behaviour is completely different from that of ordinary radiation
driven winds.
Rather than producing a steady wind with a velocity exceeding the
escape speed at the stellar surface,
the tiring  results in a complex combination of infall and outflow in 
the inner wind, with a terminal velocity in  the outer wind
that is much lower than the surface escape speed.  
The net mass flux is somewhat below the photon-tiring mass-loss limit.  
Both quantities seem to depend only weakly on the wind parameters at
the stellar surface as set by the porosity length formalism.  
(N.B. This may change for multi-dimensional simulations.)
 Less than half of the stellar luminosity escapes in the form of
radiation, even for stars that are only a small fraction above the photon tiring limit. 
For stars that are further above the photon-tiring limit, this fraction drops rapidly to less than $10\%$, 
so stars like this have an effective radiative luminosity that is less than the Eddington limit 
and could appear to be quite dim to an external observer.
  
We note in passing that the flow stagnation and a complex combination of infall
and outlfow has also proposed for the circumstellar environment of
some AGB stars \citep{s08}.
     
In future work we plan to introduce time-dependent, flux-limited, 
radiative diffusion in our simulations, first in a 1-D model, but
eventually with aims to extend this to a multi-dimensional grid.
   
We also intend to investigate the effect of stellar rotation on the
geometry of the continuum-driven wind, and how this 
might give rise to the bipolar shape seen the homonculus nebula of
$\eta~$Carinae.
     
\section*{acknowledgments}

This work was carried out with partial support from NSF grant AST-0507581. 
N.J.S. wishes to thank the support of  ISF grant 1325/06. 
We thank {J.\,MacDonald}, {N.\,Smith}, {A.\,ud-Doula} and {R.\,Townsend} for 
helpful discussions and comments.

\bsp

\appendix
\section[]{Reduced conservation equations for radiation and gas}
\label{sec-deriv}
Starting from the general mixed-frame equations of radiation
hydrodynamics as given by \cite{mm84} 
[see also \citet{s01} and \citet{st99}],
we first write the basic equations for conservation of mass, momentum, and
energy for a gas of density $\rho$, pressure $P$, 
specific internal energy 
$ e = 3P/2\rho$, 
and velocity $\vel$, 
\begin{eqnarray}
\label{gasmass}
\frac{D \rho}{D t} ~&=&~ - \rho\nabla \cdot  \vel\\
\label{gasmom}
\frac{D \vel}{Dt}~&=&~ - \frac{\nabla P}{\rho} + \grad - \ggrav  \\
\frac{De}{Dt}+ P\frac{D}{Dt} \biggl(\frac{1}{\rho}\biggr) ~&=&~
\frac{\dot{Q}}{\rho}
, 
\label{gasen}
\end{eqnarray}
where $D/Dt = \partial/\partial t + \vel \cdot \nabla$ represents  the
advective time derivative.
Here $\grad$ and $\ggrav$ are the accelerations associated
with radiative driving and gravity respectively,
while $\dot{Q}$ is the net rate per unit volume for either radiative heating
($\dot{Q}> 0$) or cooling ($\dot{Q} < 0$).

Likewise, for radiation with energy density $E$ and energy flux
$\Flux$, the associated conservation equations for radiative
momentum and energy can be written,
\begin{eqnarray}
\frac{1}{c^2} \frac{\partial \Flux}{\partial t} + \frac{1}{3}\nabla E
&=& - \rho \grad 
= - \frac{\rho \kappa \Flux}{c} 
\label{radmom}
\\
\frac{\partial E}{\partial t} + \nabla \cdot \Flux 
&=& 
- {\dot Q} -  \rho \vel \cdot \grad 
\, ,
\label{raden}
\end{eqnarray}
where eqn. (\ref{radmom}) assumes the Eddington approximation in setting 
the radiation pressure to $E/3$, 
with the extra equality giving the radiative acceleration in terms 
of the flux $\Flux$ and a grey opacity $\kappa$,
ignoring radiative drag terms of the form ${\vel}E/c$ for the 
highly subrelativistic ($\vel \ll c$) flows considered here. 
The right side of eqn. (\ref{raden}) accounts for the radiative energy
expended in heating and accelerating the gas, 
the latter representing the key ``photon tiring'' effect that is the
focus of this paper.

In practice, a further approximation here is to ignore the partial
time derivatives in both the radiative momentum and energy
equations (\ref{radmom}) and (\ref{raden}).
For the radiative flux term, this amounts to ignoring light travel
signals across the system, effectively assuming the instantaneous 
{\em propagation} of the radiative momentum in a limit of infinite light
speed, $c \rightarrow \infty$.
For the radiative energy term, it amounts to also assuming an
instantaneous {\it diffusion} of radiative energy, even compared to
hydrodynamical variations.
For a layer of optical depth $\tau$, the ratio of diffusion to
dynamical time scales scales roughly as $\tau a/c$, and since the
sound speed is $a \approx 20$~km/s, this assumption should hold for
$\tau \ll 10^{4}$.
Even for the very dense outflows considered here, this condition is
generally well satisified.

In the standard case of a static, planar, grey atmosphere in
radiative equilibrium, the right side of eqn. (\ref{raden}) vanishes, 
implying a constant radiative flux and allowing eqn. (\ref{radmom}) 
to be trivially integrated to yield a simple optical depth
scaling between radiative energy  and flux, $E \sim \tau F $.
The exchange of energy with the gas then leads to an
equilibriation of the radiative and gas temperature, resulting in
an effective ``backwarming'' of the gas at large optical depths.

An analogous backwarming can also be expected in a very dense, 
optically thick wind outlow, but the overall {\em dynamical} effect
of this is likely to be quite limited, since the associated gas
internal energy is still 
well below the gravitational binding energy.
The associated energy ratio can be cast in terms of the squared ratio
between the sound speed and escape speed, which in a wind with optical
depth $\tau$ should scale as
\begin{equation}
\frac{a^2}{\vesc^2}~=~\frac{a_\star^2}{\vesc^2} \frac{T}{T_{\rm
eff}}~\simeq~\frac{a_\star^2}{\vesc^2}\biggl[\biggl(\tau +
\frac{2}{3}\biggr)\frac{3}{4}\biggr]^{1/4}
\, ,
\end{equation}
where $a_{\star}$ is the sound speed associated with the star's
effective temperature $T_{\rm eff}$.
Since typically ${a_\star^2}/{\vesc^2} \approx 10^{-3}$,
this ratio only becomes of order unity for optical depths 
$\tau \sim 10^{12}$,
characteristic of the very center of an entire star.
It is always small for any stellar mass outflow.

Since this implies that gas internal energy is dynamically
unimportant, we ignore such optically thick backwarming effects,
and instead assume that the radiative heating term $\dot{Q}$
acts to keep the gas at a nearly {\em constant temperature},
chosen here to give a fixed sound speed 
$a = 20$~km/s that is characteristic of gas near the stellar
effective temperature $T_{\rm eff}$.
Since this implies a constant gas internal energy $e = 3 a^{2}/2$,
the energy equations for the gas (\ref{gasen}) 
and radiation (\ref{raden}) reduce to
\begin{eqnarray}
\label{gaseniso}
\frac{P}{\rho} \, \frac{D \rho}{Dt} &=& -\dot{Q} 
~=~ -  P \nabla \cdot  \vel
\\
\label{radeniso}
\nabla \cdot \Flux&=&
- \dot{Q} - \rho \vel \cdot \grad
\,  ,
\end{eqnarray}
where the latter equality in eqn. (\ref{gaseniso}) follows
from the mass conservation equation 
(\ref{gasmass}).
For the present simple case of a 1-D spherical flow with radial speed $v$
and local luminosity $L=4 \pi r^{2} F$,
these reduce to
\begin{eqnarray}
\label{gaseniso1d}
\frac{P}{\rho} \, \frac{D \rho}{Dt} &=& -\dot{Q} 
~=~ - \frac{P}{r^{2}} \, \frac{d(vr^{2})}{dr} 
\\
\label{radeniso1d}
\frac{dL}{dr} &=&
- 4 \pi r^{2} \dot{Q} - {\dot M} g_{\mathrm{rad}} 
\,  ,
\end{eqnarray}
where ${\dot M} \equiv 4 \pi \rho v r^{2}$ is the local mass flux, 
which is not necessarily constant (or even positive) in a time-dependent wind.

Note that, in this simplified formulation, the radiative momentum equation 
is superfluous, and the radiative internal energy irrelevant.
But there is still a clear global conservation of energy between the
radiation and both the work and the heating it imparts to the gas.
As such, this does provide a convenient 1-D formalism for examining the nature
of radiatively driven mass flows near the photon-tiring limit.

\label{lastpage}


\begin{thebibliography}{}
\bibitem[\protect\citeauthoryear{Begelman}{2001}]{b01}
Begelman, M.~C., 2001, \apj, 551, 897


\bibitem[\protect\citeauthoryear{CAK}{}]{cak}
Castor, J.~I., Abbott,  D.~C., \& Klein, R.~I.\ 1975, \apj, 195, 157 

\bibitem[\protect\citeauthoryear{Clark}{1996}]{c96}
Clark, D.\,A., 1996, \apj, 457,291


\bibitem[\protect\citeauthoryear{Hsu, Arons \& Klein}{1997}]{hak97} 
Hsu, J.~J.~L., Arons, J. \& Klein, R.~I., 1997, \apj, 478, 663


\bibitem[\protect\citeauthoryear{Joss, Salpeter \& Ostriker}{1973}]{jso73} 
Joss, P., Salpeter, E., and Ostriker, J., 1973, 
\apj, 181, 429

\bibitem[\protect\citeauthoryear{Mihalas \& Mihalas}{1984}]{mm84} 
Mihalas, D., \& Weibel-Mihalas, B. 1984, Foundations of Radiation Hydrodynamics,
Oxford University Press

\bibitem[\protect\citeauthoryear{Owocki, Castor \& Rybicki}{1988}]{ocr88} 
Owocki, S.\,P., Castor, J.\,I. \& Rybicki, G.\,B., 1988, \apj, 335, 914

\bibitem[\protect\citeauthoryear{Owocki, Cranmer \& Blondin}{1994}]{ocb94} 
Owocki, S.\,P., Cranmer, S.\,R. \& Blondin, J.\,M., 1994, \apj, 424, 887

\bibitem[\protect\citeauthoryear{Owocki \& Gayley}{1997}]{og97} 
Owocki, S.~P., Gayley K.~G., 1997, ASPC, 120, 121

\bibitem[\protect\citeauthoryear{Owocki \& Puls}{1999}]{op99}
Owocki, S.P., \& Puls, J. 1999,
\apj, 510, 355

\bibitem[\protect\citeauthoryear{OGS}{}]{ogs04}
Owocki, S.P., Gayley, K.G. \& Shaviv, N.J. 2004, \apj, 616, 525

\bibitem[\protect\citeauthoryear{Shaviv}{1998}]{s98} 
Shaviv, N.\,J., 1998, \apj, 494, L193

\bibitem[\protect\citeauthoryear{Shaviv}{2001a}]{s01} 
Shaviv, N.\,J., 2001a, MNRAS, 326, 126

\bibitem[\protect\citeauthoryear{Shaviv}{2001b}]{s01b} Shaviv 
N.~J., 2001b, ApJ, 549, 1093 

\bibitem[\protect\citeauthoryear{Smith}{2002}]{s02} 
Smith, N., 2002, MNRAS, 337, 1252

\bibitem[\protect\citeauthoryear{Smith \& Owocki}{2006}]{so06} 
Smith, N. \& Owocki, S.\,P., 2006, \apj, 645, L45

\bibitem[\protect\citeauthoryear{Soker}{2008}]{s08} 
Soker, N. 2008, New Astronomy, 13, 491

\bibitem[\protect\citeauthoryear{Stone \& Norman}{1992}]{sn92} 
Stone, J.\,M. \& Norman, M.L., 1992, \apjs, 80, 753

\bibitem[\protect\citeauthoryear{Spiegel}{1976}]{s76} 
Spiegel E.~A., 1976, pmas.conf, 19 


\bibitem[\protect\citeauthoryear{Spiegel \& Tao}{1999}]{st99} 
Spiegel, E., \& Tao, L. 1999,
Phys.  Rep.  311, 163

\bibitem[\protect\citeauthoryear{van Marle et al.}{2008a}]{mos08a} 
van Marle, A.\,J., Owocki,
S.\,P. \& Shaviv, N.\,J., 2008a, proceedings of: First Stars III, {Santa~Fe}, Eds.  B.
O'Shea, T. Abel, A. Heger, AIPC, 990, 250

\bibitem[\protect\citeauthoryear{Paper~1}{}]{mos08} 
van Marle, A.\,J., Owocki, S.\,P. \& Shaviv, 
N.\,J., 2008b, arXiv:0806.4536


\end{thebibliography}
\end{document}